\documentclass[aps,preprint,epsfig,prb]{revtex4}

\newcommand{\bepsilon}{\mbox{\boldmath $\epsilon$}}
\newcommand{\bv}{\mbox{\boldmath $v$}}
\newcommand{\bu}{\mbox{\boldmath $u$}}

\usepackage{amsmath}
\usepackage{graphicx}

\def\nn{\nonumber}

\begin{document}

\title{Pseudospin for Raman $D$ Band in Armchair Graphene Nanoribbons}

\author{Ken-ichi Sasaki}
\email{sasaki.kenichi@lab.ntt.co.jp}
\affiliation{NTT Basic Research Laboratories, 
Nippon Telegraph and Telephone Corporation,
3-1 Morinosato Wakamiya, Atsugi, Kanagawa 243-0198, Japan}

\author{Keiko Kato}
\affiliation{NTT Basic Research Laboratories,
Nippon Telegraph and Telephone Corporation,
3-1 Morinosato Wakamiya, Atsugi, Kanagawa 243-0198, Japan}

\author{Yasuhiro Tokura}
\affiliation{NTT Basic Research Laboratories, 
Nippon Telegraph and Telephone Corporation,
3-1 Morinosato Wakamiya, Atsugi, Kanagawa 243-0198, Japan}

\author{Satoru Suzuki}
\affiliation{NTT Basic Research Laboratories, 
Nippon Telegraph and Telephone Corporation,
3-1 Morinosato Wakamiya, Atsugi, Kanagawa 243-0198, Japan}

\author{Tetsuomi Sogawa}
\affiliation{NTT Basic Research Laboratories, 
Nippon Telegraph and Telephone Corporation,
3-1 Morinosato Wakamiya, Atsugi, Kanagawa 243-0198, Japan}

\date{\today}

\begin{abstract}
 By analytically constructing the matrix elements 
 of an electron-phonon interaction for the $D$ band 
 in the Raman spectra of armchair graphene nanoribbons, 
 we show that pseudospin and momentum conservation 
 result in (i) a $D$ band consisting of two components,
 (ii) a $D$ band Raman intensity that is enhanced only when 
 the polarizations of the incident and scattered light are 
 parallel to the armchair edge, and 
 (iii) the $D$ band softening/hardening behavior caused by the Kohn
 anomaly effect is correlated with that of the $G$ band. 
 Several experiments are mentioned that are relevant to these results.
 It is also suggested that pseudospin is independent of the boundary
 condition for the phonon mode, while momentum conservation depends on it. 
\end{abstract}


\maketitle

\section{Introduction}

Raman spectroscopy has been widely used 
for the characterization of carbon-based materials
such as graphite, graphene, nanotubes, 
and nanoribbons.~\cite{malard09}
Of the several bands that appear in the Raman spectrum, 
the $D$ band at $\sim$1350 cm$^{-1}$ is 
a mark of the presence of lattice defects.~\cite{tuinstra70} 
It is known that the formation of an electron standing wave 
due to intervalley scattering at defects 
is the key to enhancing $D$ band intensity.
Recently, the $D$ band has been examined intensively
in relation to studies of the graphene
edge.~\cite{canifmmode04sec,you08,gupta09,casiraghi09,cong10,begliarbekov10,zhang11}
In terms of symmetry,
graphene edges can be divided into two 
crystal orientation categories:
armchair and zigzag edges, 
and only an armchair edge causes intervalley scattering.
Therefore, the $D$ band is of prime importance 
when characterizing graphene edges.
In this paper, we focus on graphene nanoribbons with armchair edges
(armchair nanoribbons [ANRs]) to clarify the essential features of the $D$ band.
Several predictions that we have derived from 
pseudospin and momentum conservation for 
the matrix element of the electron-phonon interaction of the $D$ band
are useful for obtaining detailed information about
the electronic and phononic properties of armchair nanoribbons.

This paper is organized as follows.
In Sec.~\ref{sec:armwf}
we review the electronic and phononic properties of ANRs
to provide essential background about the $D$ band.
The materials described in Sec.~\ref{sec:armwf}
are needed for calculating electron-phonon matrix elements
analytically. 
In Sec.~\ref{sec:elphmat}, 
we point out several features of the $D$ band in armchair nanoribbons.
Sections~\ref{sec:dis} and~\ref{sec:con} contain our discussion and conclusion.

\section{$D$ band as first order process}\label{sec:armwf}

\begin{figure}[htbp]
 \begin{center}
  \includegraphics[scale=0.6]{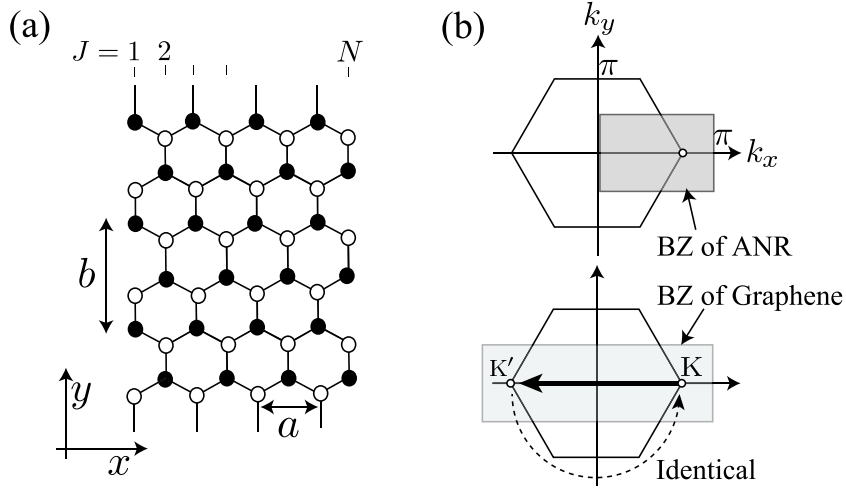}
 \end{center}
 \caption{ (a) The structure of an ANR.
 Carbon atoms are divided into A ($\bullet$) and B ($\circ$) atoms. 
 $a$ is a lattice constant [$a=2.46$ \AA] and $b\equiv \sqrt{3}a$.
 (b) Top: BZ of ANR.
 The point at $(k_x,k_y)=(2\pi/3,0)$ is the Dirac point.
 Bottom: BZ of graphene.
 The thick arrow denotes the phonon wavevector $(q_x,q_y)=(-4\pi/3,0)$
 contributing to the first order Raman process in the BZ of ANR, 
 because of zone holding. 
 }
 \label{fig:arm}
\end{figure}

The electronic energy dispersion relation of ANRs
is identical to that of graphene (without an edge).~\cite{sasaki11-armwf}
Let ${\bf k}=(k_x,k_y)$ be the wavevector of an electron, 
the energy dispersion $\varepsilon^s_{\bf k}$ is given by~\cite{wallace47}
\begin{equation}
 s\gamma 
 \sqrt{1+4\cos^2\left(\frac{k_xa}{2}\right)+4\cos\left(\frac{k_xa}{2}\right) \cos \left(\frac{k_yb}{2}\right)},
\label{eq:enearm}
\end{equation}
where $s$ ($=\pm1$) is the band index,
$\gamma$ ($=3$ eV) is the hopping integral, 
the component $k_x$ ($k_y$) of the wavevector
is perpendicular (parallel) to the edge, and 
$a$ ($b$) denotes the periodic length perpendicular (parallel) to the
edge [Fig.~\ref{fig:arm}(a)].
We use units in which $a/2=1$ and $b/2=1$, so that the wavevector and
coordinate are dimensionless quantities.

The difference between an ANR and graphene 
appears in their Brillouin zones (BZs).
The BZ of an ANR is given by $k_x \in (0,\pi)$
and $k_y \in [-\pi/2,\pi/2)$ as shown in Fig.~\ref{fig:arm}(b).
The BZ of an ANR covers only one half of the graphene's BZ.~\cite{sasaki11-armwf}
As a result, there is a single Dirac point in the BZ of the ANR, 
although there are two inequivalent Dirac points
(known as K and K$'$ points) in the BZ of graphene.~\cite{slonczewski58}
The BZ holding for ANRs can be understood more clearly 
by using the electronic standing wave in ANRs
that is written as~\cite{sasaki11-armwf} 
\begin{equation}
 \phi^s_{\bf k}(x) = \frac{1}{\sqrt{N}}e^{-ik_y(J-1)} \sin (k_x x)
  \left(
   \begin{array}{c}
    e^{-i\Theta({\bf k})} \\ s 
   \end{array}\right),
  \label{eq:bw}
\end{equation}
where $J$ $(=1,\ldots,N)$ denotes the lattice site
perpendicular to the edge [Fig.~\ref{fig:arm}(a)], and $x=J$ in units of $a/2=1$.
The upper (lower) component of Eq.~(\ref{eq:bw})
represents the amplitude at A-atom (B-atom), and
the phase $\Theta({\bf k})$
is the polar angle defined with respect
to the Dirac point: $k_y=(k_x-2\pi/3)\tan\Theta({\bf k})$.
Because $\Theta({\bf k})$ is invariant throughout the intervalley
scattering process at armchair edge,~\cite{sasaki11-armwf,sasaki10-jpsj,park11} 
$\phi^s_{-k_x,k_y}(x)=-\phi^s_{k_x,k_y}(x)$.
Thus, $\phi^s_{k_x,k_y}(x)$ and $\phi^s_{-k_x,k_y}(x)$ are identical,
that is, the reflection taking place at armchair edge identifies $(k_x,k_y)$ with $(-k_x,k_y)$,
and one half of the graphene's BZ needs to be excluded from the BZ of an
ANR to avoid a double counting.

The $x$-component of the wavevector, $k_x$,
is quantized by the boundary condition 
that is $\phi^s_{\bf k}(x)=0$ 
imposed at $x=0$ and $x=N+1$ (i.e., at fictitious edge sites), as
\begin{equation}
 k_n = \frac{n\pi}{N+1}, \ \ (n=1,\ldots,N),
 \label{eq:theta}
\end{equation}
where $n$ represents the subband index.
Note that $k_n \in (0,\pi)$ and that 
the spacing between adjacent $k_n$ is $\pi/(N+1)$.
This spacing is one half of the spacing 
obtained by imposing a periodic boundary condition on
the wave function, which shows a double density of states 
in the reduced BZ.
The double density of states in the reduced BZ is a direct consequence
of the BZ holding and 
is the principal reason why the energy gap of an ANR is different from that
of a (zigzag) nanotube obtained by rolling an ANR into a cylinder.~\cite{son06_energ}
The reduction of the BZ for ANRs is not a matter of notation but
a physical result since the energy gap is determined by the double
density of states in the reduced BZ.
Meanwhile, we assume that $k_y$ is a continuous variable.
Hereafter we abbreviate $\Theta(k_n,k_y)$ and $\phi^s_{k_n,k_y}$
by omitting $k_n$ and $k_y$ as $\Theta_n$ and $\phi^s_{n}$,
respectively.

Due to the BZ holding for ANRs,
momentum conservation is relaxed and 
a phonon with a non-zero wavevector can contribute to a first-order
Raman process. 
For example, an intervalley phonon with a wavevector $(q_x,q_y)=(-4\pi/3,0)$
can be excited without disturbing the electronic state since 
the final electronic state (K$'$)
is identical to the initial state (K),
as shown by the dashed arrow in the lower inset to Fig.~\ref{fig:arm}(b).
In graphene without an edge,
it is considered that for a photo-excited electron at $(k_x,k_y)$,
only the phonon at the $\Gamma$ point (${\bf q}=0$)
contributes to the first order Raman process 
due to momentum conservation.~\cite{malard09}
However, in ANRs, a phonon with 
the wavevector $(q_x,q_y)=(-2k_x,0)$ [mod $\pi$]
contributes to the first order Raman process.~\footnote{
If a phonon with non-zero $q_y$ [$(q_x,q_y)=(-2k_x,q_y)$] is involved, 
the final electronic state is distinct from the initial state.
Then the excitation of the phonon does not satisfy 
the condition of the first order process, although the phonon may be involved 
in a second order process such as the 2$D$ band (overtone of $D$ band).}
As we will show later with the more mathematically rigorous method,
the $D$ band is represented as a first order, single resonance
process in the reduced BZ, whereas it is commonly recognized
as a second order, double resonance process in the BZ of
graphene.~\cite{thomsen00,saito03}

\begin{figure}[htbp]
 \begin{center}
  \includegraphics[scale=0.5]{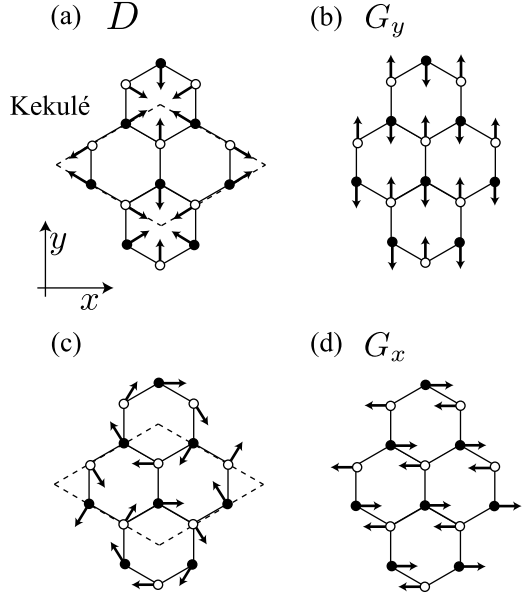}
 \end{center}
 \caption{The displacement vectors of four phonon modes.
 (a) Kekul\'e distortion $-\bu_{2\pi/3}(x)$, 
 (b) $-\bu_{0}$ (or $G_y$) representing an optical phonon at the $\Gamma$ point, 
 (c) $\bv_{2\pi/3}(x)$, and (d) $\bv_{0}$ (or $G_x$).
 }
 \label{fig:dis}
\end{figure}

Vibrational displacement vectors for 
phonons with ${\bf q}=(q,0)$
in ANRs can be explicitly constructed.
First, we define the acoustic and optical branches
\begin{align}
 \sin(q x)
 \begin{pmatrix}
  {\bf e}_i \cr {\bf e}_i
 \end{pmatrix}, \ \ 
 \cos(q x)
 \begin{pmatrix}
  {\bf e}_i \cr -{\bf e}_i
 \end{pmatrix},
 \label{eq:mode}
\end{align}
as the basic vibrational modes in ANRs.
In Eq.~(\ref{eq:mode}),
the upper (lower) component represents the vibrational direction
of an A-atom (B-atom), where ${\bf e}_i$ ($i=x,y$) is the unit vector
parallel to the $i$-axis.
The function $\sin(q x)$ or $\cos(q x)$
represents the vibrational amplitude of the atom at $x$.~\footnote{
Note that the BZ for a phonon
is given by one half of the graphene's BZ, $q \in [0,\pi]$,
as well as the ANR's BZ for an electron.}
The amplitude may be normalized by multiplying 
each mode with the proper amplitude $u$.
For the optical modes, $u$ is nearly constant at about $0.01$\AA,
independent of the value of $q$.
For the acoustic mode, 
$u$ depends sensitively on
the energy, which is a function of $q$ and temperature.
We generally have different amplitudes 
for acoustic and optical modes.
However, at the point $q = 2\pi/3$
the energies of the acoustic modes become comparable to that of the
optical modes, so the amplitudes of the acoustic and optical modes
are approximately the same.~\footnote{
The following analysis is valid even though
we take into account the difference of the amplitudes of the acoustic
and optical modes. 
See Appendix~\ref{app:kekule} for detailed discussion of this point.}
Then normal modes would be given by the sum of the acoustic
and optical modes. 
From the basic vibrational modes given 
in Eq.~(\ref{eq:mode}),
we construct new modes as 
\begin{align}
 {\bu}_{q}(x) = \sin(q x) 
 \begin{pmatrix}
  {\bf e}_x \cr {\bf e}_x
 \end{pmatrix}
 + \cos(q x) 
 \begin{pmatrix}
  {\bf e}_y \cr -{\bf e}_y
 \end{pmatrix}.
\label{eq:umodes}
\end{align}
This mode is a superposition of 
the acoustic mode along the $x$-axis (LA)
and the optical mode along the $y$-axis (TO).
The mode ${\bu}_{q}(x)$ with $q = 2\pi/3$
corresponds to the Kekul\'e distortion shown in Fig.~\ref{fig:dis}(a).
In Fig.~\ref{fig:dis}(a), it is clear that 
the Kekul\'e distortion does not change the bond angle, and that 
the mode is composed only of a bond stretching motion.~\cite{tuinstra70}
The mode ${\bu}_{q}(x)$ with $q=0$
can be used to represent the 
$\Gamma$ point optical mode shown in Fig.~\ref{fig:dis}(b) 
because the acoustic component in Eq.~(\ref{eq:umodes})
disappears.
We define another combination of acoustic and optical branches as 
\begin{align}
 {\bv}_{q}(x) = \cos(q x) 
 \begin{pmatrix}
  {\bf e}_x \cr -{\bf e}_x
 \end{pmatrix}
 - \sin(q x) 
 \begin{pmatrix}
  {\bf e}_y \cr {\bf e}_y
 \end{pmatrix}.
\label{eq:vmodes}
\end{align}
This mode is a superposition of 
the optical mode along the $x$-axis (LO)
and the acoustic mode along the $y$-axis (TA).
The mode ${\bv}_{q}(x)$
is given by a $-90^\circ$ rotation of the mode ${\bu}_{q}(x)$:
$\bv_{q}(x)=R(-\pi/2){\bu}_{q}(x)$.
Thus, the mode ${\bv}_{q}(x)$ with $q = 2\pi/3$
is realized simply by a change in the bond angle,
as shown in~\ref{fig:dis}(c), and 
the mode ${\bv}_{q}(x)$ with $q=0$ is 
the $\Gamma$ point optical mode shown in Fig.~\ref{fig:dis}(d).

According to Tuinstra and Koenig,~\cite{tuinstra70}
the $D$ band originates from the Kekul\'e distortion,
while the $G$ band is composed of the two optical components
$\bu_{0}$ and $\bv_{0}$.
The vibrational energy of the Kekul\'e distortion
($\omega^2=3K/M$ where $M$ is the mass of the carbon
atoms~\cite{yoshimori56})
is highest among the in-plane phonons at the Dirac point,
while that for ${\bv}_{2\pi/3}$ ($\omega^2=18H/M$) is the
lowest, because the bond stretching force constant ($K$)
is much larger than the force constant 
for the deformation of the angle ($H$).~\cite{tuinstra70}
The $\bu_{0}$ and $\bv_{0}$ energies
are given by $\omega^2=3K/M+18H/M$, showing that the $G$ band
originates from both bond stretching and bond angle
change.~\cite{tuinstra70}
Hereafter, we refer to the component of the $G$ band
that originates from $\bu_{0}$ 
($\bv_{0}$) as $G_y$ ($G_x$) band for clarity.

\section{Electron-phonon interaction}\label{sec:elphmat}

First, we consider the electron-phonon interaction 
for the vector ${\bu}_q(x)$.
The bond stretching induced by a displacement vector 
gives rise to two distinct effects:
a change in the nearest-neighbor hopping integral and 
the generation of a (deformation) potential at each atom.
The former is known as an off-site interaction
and the latter as an on-site interaction.
Therefore, the electron-phonon interaction
consists of off-site ($H_{\rm off}$) and on-site ($H_{\rm on}$)
components.
The matrix elements for each component 
are given by (see Appendix~\ref{app:ep} for derivation)
\begin{align}
 \langle \phi^{s'}_m| H_{\rm off}({\bu}_q) | \phi^{s}_n \rangle 
 =  \Bigg\{ &
 -\left( \frac{2}{N} \sum_{J=1}^N
 \sin(k_m J)\cos(q J) \sin(k_n J)\right) 
 \left[ 2+ 2\sin\left(\frac{q}{2}+\frac{\pi}{6}\right) \cos\left(\frac{q}{2}
 \right) \right] 
 \nn \\
 &- \left( \frac{2}{N} \sum_{J=1}^N 
 \sin(k_m J) \sin(q J) \cos(k_n J) \right)
 \left[4  \sin\left(\frac{q}{2}+\frac{\pi}{6}\right) 
 \sin\left(\frac{q}{2} \right) \sin (k_n) \right] \Bigg\}
 \langle \sigma_x \rangle_{mn}^{s's},
 \label{eq:Mu}
\end{align}
and
\begin{align}
 \langle \phi^{s'}_m| H_{\rm on}({\bu}_q) | \phi^{s}_n \rangle 
 = \left( \frac{2}{N} \sum_{J=1}^N
 \sin(k_m J)\cos(q J) \sin(k_n J)\right)
 \left[ 4 \sin\left(\frac{q}{2}
 \right) \cos\left(\frac{q}{2}+\frac{\pi}{6}\right)  \right] 
 \langle \sigma_0 \rangle_{mn}^{s's},
 \label{eq:Mu_on}
\end{align}
where the matrix element of the Pauli matrix $\sigma_i$
($i=x,y,z,0$) is defined as
\begin{align}
 \langle \sigma_{i} \rangle^{s's}_{mn} \equiv  
 \frac{1}{2}
 \begin{pmatrix}
  e^{i\Theta_m} & s' 
 \end{pmatrix}
 \sigma_{i}
 \begin{pmatrix}
  e^{-i\Theta_n} \cr s
 \end{pmatrix},
\label{eq:sigmaexp}
\end{align}
$\sigma_0$ is the 2$\times$2 identity matrix and
\begin{align}
 \sigma_x=
 \begin{pmatrix}
  0 & 1 \cr 1 & 0
 \end{pmatrix},
 \sigma_y=
 \begin{pmatrix}
  0 & -i \cr i & 0
 \end{pmatrix},
 \sigma_z=
 \begin{pmatrix}
  1 & 0 \cr 0 & -1
 \end{pmatrix}.
\end{align}
The right-hand side of the off-site component Eq.~(\ref{eq:Mu})
is proportional to the pseudospin $\langle \sigma_x \rangle_{mn}^{s's}$,
while the on-site component depends on $\langle \sigma_0 \rangle_{mn}^{s's}$.
The summations over $J$ in Eq.~(\ref{eq:Mu}) and Eq.~(\ref{eq:Mu_on}),
i.e. $(2/N \sum_{J=1}^N \ldots)$, do not vanish only when 
$q$ are specific values that satisfy momentum conservation.
So, the matrix element is determined 
by two factors: pseudospin and momentum conservation.

Let us focus on the first order Raman process, 
$\phi_n^s \to \phi_n^s$ (i.e., $k_m=k_n$ and $s'=s$), 
that is, the diagonal matrix element in Eq.~(\ref{eq:Mu}).
By executing the summation over $J$ for momentum conservation,
we see that the matrix element can be non-zero if
the phonon's wavevector $q$ satisfies 
(i) $q=0$ or (ii) $q=2\pi-2k_n$.
The case (i) corresponds to the $\Gamma$ point optical phonon,
which is relevant to $G_y$,
while the case (ii) is relevant to the $D$ band since 
$q \simeq 2\pi/3$ is satisfied 
for the electronic states near the Dirac point 
($k_n \simeq 2\pi/3$).

\begin{figure}[htbp]
 \begin{center}
  \includegraphics[scale=0.5]{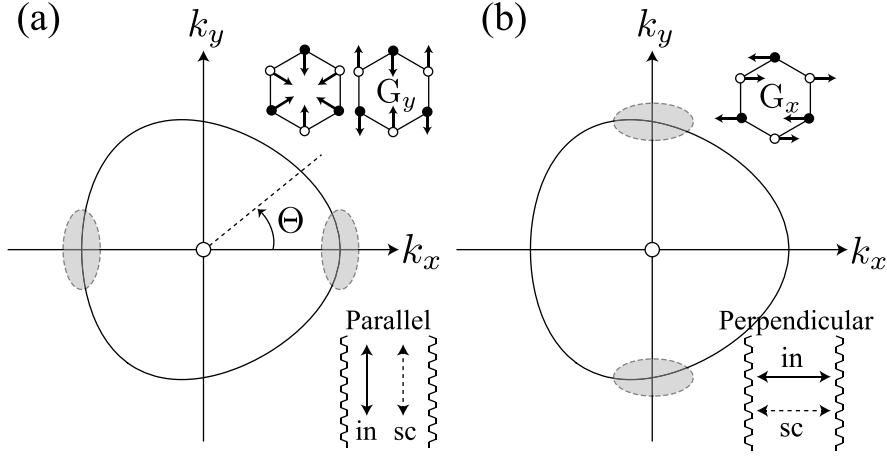}
 \end{center}
 \caption{Energy contour circle near the Dirac point.
 (a) 
 The electrons at the two shaded regions on the energy
 contour circle ($\Theta\simeq 0$ or $\pi$), 
 have the strongest electron-phonon matrix element 
 for emitting the $D$ and $G_y$ bands. 
 These electrons can be selectively excited by the light polarization
 setup: ``in'' (``sc'') denotes the polarization of incident
 (scattered) light.
 (b) 
 The electrons in the two shaded regions ($\Theta\simeq \pi/2$ or $-\pi/2$),
 have the strongest electron-phonon matrix element 
 for emitting the $G_x$ band. 
 }
 \label{fig:pol}
\end{figure}

For the $D$ band,
by setting $q \simeq 2\pi/3$ in Eq.~(\ref{eq:Mu}), we obtain
\begin{align}
 \langle \phi^s_n| H_{\rm off}({\bu}_{q \simeq 2\pi/3}) | \phi^s_n \rangle
 \simeq 3 \langle \sigma_x \rangle_{nn}^{ss}.
 \label{eq:epD}
\end{align}
From Eq.~(\ref{eq:Mu_on}), 
the on-site component almost vanishes 
$\langle \phi^s_n| H_{\rm on}({\bu}_{q \simeq 2\pi/3}) | \phi^s_n \rangle \simeq 0$,
so that only the off-site component contributes to the matrix element.
The off-site component is proportional to 
$\langle \sigma_x \rangle_{nn}^{ss}$, which is 
$s\cos\Theta_n$ from Eq.~(\ref{eq:sigmaexp}).
The transition probability $\sim \cos^2\Theta_n$
takes its maximum (minimum) value when $\Theta_n =0$ or $\pi$
($\Theta_n =\pi/2$ or $-\pi/2$).
Thus, the electrons at $\Theta_n =0$ or $\pi$
contribute selectively to the $D$ band Raman intensity, while
the phonon emission from 
the electrons at $\Theta_n =\pi/2$ or $-\pi/2$ is negligible
[Fig.~\ref{fig:pol}(a)].
From the wave function Eq.~(\ref{eq:bw}),
the photo-excited electron at $\Theta=0$ ($\pi$) 
corresponds to the bonding (antibonding) orbital.
This result suggests that the bonding and antibonding orbitals 
couple strongly to the bond stretching motion, which
can be also understood by diagonalizing the 2$\times$2 matrix
$H=\gamma \sigma_x$ for a two-atom system.
The eigenstates of $H$ are bonding or antibonding orbitals.
Since the bond stretching motion induces a change in $\gamma$ as
$\gamma+\delta \gamma$, the off-site electron-phonon interaction is
given by $H_{\rm off}=\delta \gamma \sigma_x$, whose expectation values
become its maximum for the bonding or antibonding orbitals.
Since the $D$ band consists only of the bond stretching motion,
the electrons at $\Theta_n =0$ and $\pi$ 
couple strongly to the $D$ band phonon.

For the $G_y$ band, 
by setting $q=0$ in Eq.~(\ref{eq:Mu}), we obtain 
\begin{align}
 \langle \phi^s_n| H_{\rm off}(\bu_{0}) | \phi^s_n \rangle
 =-3 \langle \sigma_x \rangle_{nn}^{ss}.
 \label{eq:uyg}
\end{align}
Because there is no contribution from the on-site component
[$\langle \phi^s_n| H_{\rm on}(\bu_{0}) | \phi^s_n \rangle =0$ from Eq.~(\ref{eq:Mu_on})],
the transition amplitude is determined by the off-site component.
It is easy to see that only
the electrons at $\Theta_n =0$ or $\pi$
contribute to the $G_y$ band intensity.

Comparing Eq.~(\ref{eq:epD}) with Eq.~(\ref{eq:uyg}),
we see that the electron-phonon matrix elements 
for the $D$ and $G_y$ bands are the same 
(apart from the unimportant sign change),
although the phonon's wave vectors are distinct.
A direct consequence of this fact 
is that if the $D$ band is activated,
the $G_y$ band should be activated in the same manner.
For example, 
the $D$ band intensity at armchair edge becomes comparable to the $G$
band intensity.
In addition,
due to the fact that both the electron-phonon matrix elements are
proportional to $\langle \sigma_x \rangle_{nn}^{ss}$,
the $D$ and $G_y$ band Raman intensities have the same polarization dependence:
the Raman intensities of these phonon modes
should be enhanced
when the polarization of the incident light
is parallel to the armchair edge.
This polarization dependence of the $D$ band has been confirmed 
in many experiments.~\cite{canifmmode04sec,you08,gupta09,casiraghi09}
Several experimental groups have observed a correlation between the $D$ and $G$
bands,~\cite{canifmmode04,cong10,begliarbekov10,zhang11,sasaki10-jpsj}
suggesting that the $G_x$ Raman intensity is suppressed at the armchair edge.
Another example is the correlation between the frequency
hardening/softening behavior of the $D$ band due to doping (Kohn anomaly
effect) and that of the $G_y$ band.
These features of the $D$ and $G$ bands 
are further explored in the following subsections.

Next, we examine the electron-phonon interaction
caused by the vector
${\bv}_{q}(x)$ defined in Eq.~(\ref{eq:vmodes}).
The corresponding electron-phonon matrix element 
is given by 
\begin{align}
 \langle \phi^{s'}_m|  H_{\rm off}({\bv}_q) | \phi^s_n \rangle
 = \Bigg\{ 
 &-\left(
 \frac{2}{N} \sum_{J=1}^N 
 \sin(k_m J) \sin(q J) \sin(k_n J) \right) \left[
 2 \sin\left(\frac{q}{2} \right)
 \cos\left(\frac{q}{2}+\frac{\pi}{6}\right)  
 \right] \nn \\
 &- 
 \left(\frac{2}{N} \sum_{J=1}^N 
 \sin(k_m J) \cos(q J)\cos(k_n J) \right)
 \left[ 4 \cos\left(\frac{q}{2} \right) 
 \cos\left(\frac{q}{2}+\frac{\pi}{6}\right) 
 \sin(k_n) \right] \Bigg\}
 i \langle \sigma_y \rangle^{s's}_{mn}, 
 \label{eq:Mv}
\end{align}
and
\begin{align}
 \langle \phi^{s'}_m| H_{\rm on}({\bv}_q) | \phi^s_n \rangle 
 = \left(
 \frac{2}{N} \sum_{J=1}^N 
 \sin(k_m J) \sin(q J) \sin(k_n J) \right) \left[
 4 \sin\left(\frac{q}{2} \right)
 \cos\left(\frac{q}{2}+\frac{\pi}{6}\right)  
 \right] \langle \sigma_z \rangle^{s's}_{mn}.
 \label{eq:Mv_on}
\end{align}
A notable difference between 
the matrix elements for ${\bu}_q(x)$ and ${\bv}_q(x)$
is that the off-site components are proportional to $\sigma_x$ and
$\sigma_y$, respectively.
Moreover, the on-site interaction
induced by ${\bu}_q(x)$ is symmetric about two sublattices, 
while that induced by ${\bv}_q(x)$ is antisymmetric, 
as is evident from the matrices $\sigma_0$ and
$\sigma_z$ in Eq.~(\ref{eq:Mu_on}) and Eq.~(\ref{eq:Mv_on}).

From Eq.~(\ref{eq:Mv_on})
the on-site component is strongly suppressed 
for the phonon satisfying $q \simeq 0$ or 
$q \simeq 2\pi/3$, and it is negligible.
For $q \simeq 2\pi/3$,
the off-site component is also negligible as
$\langle \phi^s_m|H_{\rm off}({\bv}_{q \simeq 2\pi/3})|\phi^s_n
\rangle \simeq 0$
because $\cos(q/2+\pi/6)\simeq 0$ holds.
Thus, the mode ${\bv}_{q \simeq 2\pi/3}(x)$ 
does not appear in the Raman spectrum.
Since ${\bv}_{q \simeq 2\pi/3}(x)$
consists only of a change in bond angle and does not contain any bond
stretching motion, 
the vanishing matrix elements are reasonably understood to show that
only the bond stretching motion gives rise to the electron-phonon interaction.
For $q \simeq 0$, on the other hand, the off-site component 
can be non-zero as 
\begin{align}
 \langle \phi^{s'}_m| H_{\rm off}({\bv}_{q \simeq 0}) | \phi^s_n \rangle
 \simeq 
 -\frac{1}{\pi}\left(\frac{1}{m-n+p}
 +\frac{1}{m-n-p} \right) 
 3 i \langle \sigma_y \rangle^{s's}_{mn},
\label{eq:elphGx}
\end{align}
where $m-n\pm p$ ($\equiv \Delta n$) is an odd number
and integer $p$ is defined through $q=p \pi/(N+1)$. 
When $m-n\pm p$ is an even number, 
the off-site element vanishes.
The factor $(1/\pi \Delta n)$ 
on the right-hand side of Eq.~(\ref{eq:elphGx}),
originates from momentum conservation,
and a small change in the wave number ($m-n\ne 0$)
is needed to excite this $G_x$ band.
A large change in the wave number is not important because the
matrix element is suppressed by the inverse of $\Delta n$.
By setting $s'=s$ and $m\simeq n$ in Eq.~(\ref{eq:sigmaexp}), we get
$\langle \sigma_y \rangle_{mn}^{ss}\simeq s\sin\Theta_n$.
The transition amplitude 
takes its maximum value when $\Theta_n =\pi/2$ or $-\pi/2$.
Thus, the electrons at $\Theta_n =\pi/2$ or $-\pi/2$
contribute selectively to the $G_x$ band Raman intensity, while 
the phonon emission from 
the electrons at $\Theta_n =0$ or $\pi$ is negligible
[Fig.~\ref{fig:pol}(b)].
In the following sections, we point out several consequences 
that can be derived from the electron-phonon matrix elements.

\subsection{$D$ band splitting}\label{ssec:tp}

The fact that there are the two electronic states with bonding or
antibonding orbitals ($\Theta=0$ or $\pi$)
from which the $D$ band phonon is emitted 
results in the splitting of the $D$ band if we take account of
the trigonal warping effect.
The splitting width increases with
increasing incident laser energy $E_L$ as
\begin{align}
 \Delta \omega_D = 25 \left(\frac{E_L}{\gamma}\right)^2 \ 
 [{\rm cm}^{-1}],
 \label{eq:disp}
\end{align}
which is about 15 cm$^{-1}$ when $E_L=2.3$ eV.
This formula is derived as follows.

\begin{figure}[htbp]
 \begin{center}
  \includegraphics[scale=0.5]{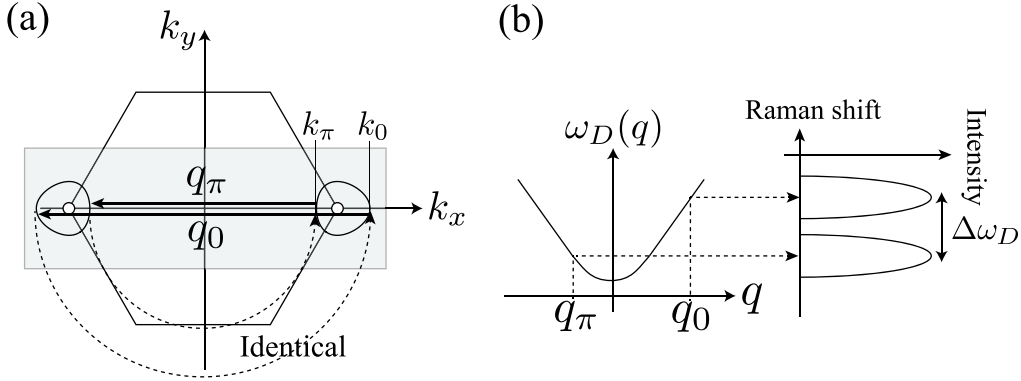}
 \end{center}
 \caption{(a) Two principal wavevectors of the phonons ($q_0$ and $q_\pi$)
 contributing to the $D$ band.
 (b) The phonon dispersion relation near the Dirac point.
 The energy difference between $\omega_D(q_0)$ and $\omega_D(q_\pi)$
 appears as two peaks in the $D$ band.
 }
 \label{fig:split}
\end{figure}

Since the probability that the photo-excited electron at
$\Theta$ emits the Kekul\'e mode is proportional to $\cos^2 \Theta$ [Eq.~(\ref{eq:epD})],
the photo-excited electrons satisfying 
$\Theta \simeq 0$ or $\Theta \simeq \pi$,
indicated by the shaded regions shown in Fig.~\ref{fig:pol}(a), 
contribute efficiently to the $D$ band intensity.
Thus, two phonon modes that originate from the two regions 
in the photo-excited electron contribute to the $D$ band.
We denote $k_x$ at $\Theta=0$ ($\Theta=\pi$) as $k_0$ ($k_\pi$)
[see Fig.~\ref{fig:split}(a)], where
$k_0$ and $k_\pi$ are given by solving the equation
$E_L/2=\gamma|1+2\cos(k_x)|$ as 
\begin{align}
\begin{split}
 & k_{0}-\frac{2\pi}{3}
 =\frac{\sqrt{3}}{2} \left(
 \frac{E_L}{3\gamma}\right)
 +\frac{\sqrt{3}}{8} \left(\frac{E_L}{3\gamma}\right)^2 +\cdots, \\
 & \frac{2\pi}{3}-k_{\pi}=
 \frac{\sqrt{3}}{2} \left(
 \frac{E_L}{3\gamma}\right)
 -\frac{\sqrt{3}}{8} \left(\frac{E_L}{3\gamma}\right)^2 +\cdots.
\end{split}
\label{eq:thetae}
\end{align}
The $\pm$ signs in front of $\sqrt{3}/8 (E_L/3\gamma)^2$
represent the trigonal warping effect meaning that 
the energy contour circle is anisotropic
as shown in Figs.~\ref{fig:pol} and~\ref{fig:split}(a).
The corresponding phonon's wavevector
is given by $q_{0/\pi}=2\pi-2k_{0/\pi}$
[Fig.~\ref{fig:split}(a)].
Then we find that the difference between 
the phonon's wavevectors $|q_0-2\pi/3|$ 
(relative to the Dirac point) and $|q_\pi-2\pi/3|$ is well approximated
by
\begin{align}
 \Delta q = \frac{\sqrt{3}}{2} \left(\frac{E_L}{3\gamma}\right)^2.
 \label{eq:dtheta_p}
\end{align}
For simplicity, let us assume that 
the phonon dispersion relation for the $D$ band 
is isotropic about the Dirac point: $\omega_D$ is a function of 
$|q-2\pi/3|$ as $\omega_D(|q-2\pi/3|)$.
From Fig.~\ref{fig:split}(b) it is clear that the two phonon modes with
$q_0$ and $q_\pi$ have different phonon energies, which results in
the double peak structure of the $D$ band.

The difference between the energy for the phonon mode with
$q_0$ and that for $q_\pi$ is approximated by
\begin{align}
 \Delta \omega_D = \frac{\partial \omega_D}{\partial q}
 \Delta q,
 \label{eq:dome_D}
\end{align}
where $\partial \omega_D/\partial q$
is the slope of the phonon energy dispersion.
The value of the slope can be estimated from the dispersive behavior of
the $D$ band: the $D$ band frequency increases as the laser energy $E_L$ increases. 
For example, Matthews {\it et al}.~\cite{matthews99} reported that 
$\partial \omega_D/\partial E_L$ 
for graphite is about 50 cm$^{-1}/$eV.
A similar value has been obtained for single layer
graphene.~\cite{gupta09,casiraghi09}
We interpret the dispersive behavior that occurs 
as a result of the dispersion relation of the phonon mode.~\cite{matthews99} 
Then we have
\begin{align}
 \frac{\partial \omega_D}{\partial E_L} = 
 \frac{\partial q}{\partial E_L}
 \frac{\partial \omega_D}{\partial q}.
 \label{eq:disperivep}
\end{align}
By combining $E_L/2=\gamma|1+2\cos(k)|$ and $q=2\pi-2k$,
we have $|\partial E_L/\partial q| \approx \sqrt{3}\gamma$
for the states near the Dirac point. 
Putting this result in Eq.~(\ref{eq:disperivep}),
we have $\partial \omega_D/\partial q=\sqrt{3}\gamma (\partial \omega_D/\partial E_L)$.
Then we obtain Eq.~(\ref{eq:disp})
by combining Eq.~(\ref{eq:dtheta_p}) and Eq.~(\ref{eq:dome_D}).
Actual splitting can be smaller (or even larger) than Eq.~(\ref{eq:disp})
due to several factors,
such as (i) 
the phonon dispersion relation for the $D$ band 
not being exactly isotropic around the Dirac point,
(ii) $\Theta$ is not exactly limited by $0$ or $\pi$.
The conditions for observing the splitting are 
discussed later.

\subsection{Light polarization dependence}\label{ssec:lpd}

Pseudospin is the key to the characteristic polarization
dependence of the $D$ and $G$ band intensities.
Below we show that the Raman intensities are enhanced only when 
the polarizations of the incident and scattered light are both parallel to
the armchair edge. That is,
the $D$ band intensity is well approximated by
\begin{align}
 I_{\rm D}(\Theta_{\rm in},\Theta_{\rm sc}) 
 \simeq I \cos^2 (\Theta_{\rm in}) \cos^2 (\Theta_{\rm sc}),
\end{align}
where $\Theta_{\rm in}$ ($\Theta_{\rm sc}$)
the angle between the orientation of the armchair edge 
and the incident (scattered) light polarization.~\cite{sasaki10_pola}
Without a polarizer for the scattered light,
we obtain $I_{\rm D}(\Theta_{\rm in}) \simeq I\pi \cos^2 (\Theta_{\rm in})$
by integrating $I_{\rm D}(\Theta_{\rm in},\Theta_{\rm sc})$ over $\Theta_{\rm sc}$.

To discuss light polarization dependence,
let us review the interband optical matrix elements 
obtained in a previous paper.~\cite{sasaki11-dc}
The electron-light interaction is written as 
$H(\bepsilon)=-e \bv\cdot {\bf A}$,
where $-e$ is the electron charge,
$\bv=(v_x,v_y)$ is the velocity operator,
and ${\bf A}=A \bepsilon e^{-i\omega t}$ is a spatially
uniform vector potential.
Here, $\hbar \omega$ corresponds to the 
incident light energy and $\bepsilon$ denotes the polarization.
The calculated matrix elements are given in units of $-ev_{\rm F}A$
by 
\begin{align}
 & \langle \phi^c_m | H(\epsilon_x) | \phi^v_n \rangle 
 = \begin{cases}
  \displaystyle 0 & \text{$m-n \in$ even}, \\
  \displaystyle - \frac{2}{\pi} \frac{\langle \sigma_y \rangle_{mn}^{cc}}{m-n}  & \text{$m-n \in$ odd},
 \end{cases}
\label{eq:interx} \\
 & \langle \phi^c_m | H(\epsilon_y) | \phi^v_n \rangle 
 = i \delta_{mn} \langle \sigma_x \rangle_{mn}^{cc}.
\label{eq:intery}
\end{align}
The Kronecker delta $\delta_{mn}$
in Eq.~(\ref{eq:intery})
shows that the $y$-polarized light ($\epsilon_y$: parallel to the
armchair edge) results in a direct interband transition ($m=n$). 
The transition amplitude depends on 
the diagonal matrix element
of the pseudospin $\langle \sigma_x \rangle_{nn}^{cc}$.
As we have seen,
$|\langle \sigma_x \rangle^{cc}_{nn}|$ takes its maximum value
for $\Theta_n=0$ or $\pi$.
On the other hand, 
$|\langle \sigma_x \rangle_{nn}^{cc}|$ vanishes for 
$\Theta_n=\pm \pi/2$.
Therefore, only the electrons at $\Theta_n=0$ or $\pi$
are selectively excited by the $y$-polarized light.
From Eq.~(\ref{eq:interx}),
the $x$-polarized light ($\epsilon_x$: perpendicular to the edge)
results in an indirect interband transition
($m-n=\pm1,\pm3,\ldots$), and 
the optical transition probability 
is reduced by the factor of $(2/\pi)^2(m-n)^{-2}$, 
independent of the pseudospin $\langle \sigma_y \rangle_{mn}^{cc}$.

First, we consider a parallel configuration:
both the incident and scattered lights are $y$-polarized
[see Fig.~\ref{fig:pol}(a)].
Since the absorption and emission of $y$-polarized light 
occur through direct transitions $\phi_n^v \leftrightarrow \phi_n^c$, 
the phonon mode with $q=0$ or $q=2\pi-2k_n$
fulfils the momentum conservation of the first order Raman process.
Thus, the $D$ and $G_y$ bands can be detected 
by the parallel configuration.
The $G_x$ band is suppressed 
because its activation requires a non-zero shift in the electron wave
number, and the momentum deficit cannot be compensated by 
direct optical transitions.
More importantly, 
pseudospin enhances (suppresses) the Raman
intensities of the $D$ and $G_y$ bands  ($G_x$ band)
because the probability amplitude of exciting the $D$ and $G_y$ bands
($G_x$ band) is proportional to $\cos\Theta_n$ ($\sin\Theta_n$).
The function $\cos\Theta_n$ agrees with 
the optical matrix element that is also proportional to $\cos\Theta_n$,
while the function $\sin\Theta_n$ is out-of-phase. 
The probability that 
the electron at $\Theta_n$ completes the Raman process
by emitting the D or $G_y$ band ($G_x$ band)
is proportional to $\cos^6\Theta_n$ ($\cos^4\Theta_n \sin^2\Theta_n$).
We estimate the reduction in the intensity of the $G_x$ band caused by
the disagreement of the pseudospin to be 1/5 by using the ratio of 
$\oint \cos^4\Theta_n \sin^2\Theta_n d\Theta_n$ ($=\pi/8$) to 
$\oint \cos^6\Theta_n d\Theta_n$ ($=5\pi/8$).
The $G_x$ band intensity is further reduced by the momentum deficit.

Next, we examine a crossed configuration:
the incident light is $y$-polarized, while the scattered light is
$x$-polarized. 
Since $x$($y$)-polarized light results in an indirect (direct) optical
transition, momentum mismatch takes place:
the electron cannot return to its original state in the valence band
unless a phonon with a specific momentum is involved in the Raman process.
The intensity is suppressed 
by a factor of $(2/\pi)^2$ because of 
the reduction factor in the optical matrix element for the $x$-polarized
light. 
Furthermore,
momentum conservation for the phonon mode reduces the
emission probability by the factor of $(2/\pi)^2$ for the $G_x$ band 
[or $(1/2)^2$ for the $D$ and $G_y$ bands], 
so that the intensity is suppressed by a factor of $(2/\pi)^4$
($\simeq 0.16$).
Moreover, the pseudospin for incident light does not match 
that for scattered light, and 
the Raman intensity is suppressed by a factor of $1/5$
in the crossed configuration.
Thus, the $D$ and $G_y$ bands intensities are suppressed by a few percent
compared with the intensities in the parallel configuration.

Finally, we study a perpendicular configuration:
both the incident and scattered lights are $x$-polarized
[see Fig.~\ref{fig:pol}(b)].
The Raman intensities of the $D$, $G_y$, and $G_x$ bands 
for this configuration are suppressed 
by $(2/\pi)^4$ at least,
due to the reduction factor for the optical matrix elements.
The pseudospin for the $x$-polarized light agrees only with 
that for the $G_x$ band.
Thus, the $D$ and $G_y$ bands are not detectable 
in the perpendicular configuration because of
the further reduction by a factor of $1/5$.

When the polarization of the incident light is perpendicular to the
armchair edge, it is common for the $D$ band to show a residual
intensity.~\cite{canifmmode04sec,casiraghi09,gupta09,cong10}
The ratio of the minimum $D$ band intensity (observed when the
incident light polarization is perpendicular to the edge) to the
maximum intensity (observed when the incident light polarization is 
parallel to the edge) is 0.12 $\sim$ 0.2 depending on the experiments.
Several authors have suggested that the presence of the residual $D$ band
intensity is correlated with the irregularities that consist of armchair
segments having angles of $\pm 60^\circ$ with respect to the armchair
edge.~\cite{canifmmode04sec,casiraghi09,xu11}
Because their arguments assume that the irregularities contribute
independently to the total $D$ band intensity as 
$I_{\rm D}(\Theta_{\rm in}) \simeq a \cos^2 (\Theta_{\rm in}) +
b\{\cos^2(\Theta_{\rm in}+60^\circ)+\cos^2(\Theta_{\rm in}-60^\circ)\}$,
there is a possibility that 
this assumption is not plausible when the electron wave function
extends along the armchair edge.
We provide a different interpretation of the residual $D$ band intensity.
When the momentum conservation that we have used 
is not physically legitimate because of the
boundary condition of the
phonon mode (see Appendix~\ref{app:ep} for a relevant discussion),  
we can only rely on the pseudospin.
Then we may expect the residual $D$ band intensity to be about $0.2$
($1/5$).
Even for an armchair edge without irregularity,
the $D$ band exhibits a residual intensity depending on the boundary
condition of the phonon mode.

\begin{figure}[htbp]
 \begin{center}
  \includegraphics[scale=0.4]{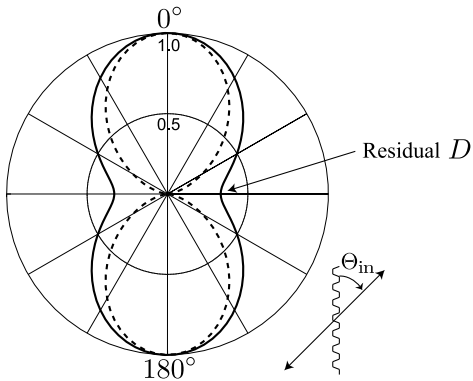}
 \end{center}
 \caption{The polar plot for the $D$ band intensity.
 The parameters for the solid curve are $a=1$, $b=1/3$, and $c=0$,
 while those for the dashed curve are $a=1$ and $b=c=0$.
 }
 \label{fig:polarp}
\end{figure}

Without a polarizer for the scattered light,
the $D$ band polarization dependence is expressed as
\begin{align}
 I_{\rm D}(\Theta_{\rm in})
 = a \cos^2 (\Theta_{\rm in}) + b \sin^2 (\Theta_{\rm in}) +c,
 \label{eq:ID}
\end{align}
where $a$, $b$, and $c$ are parameters.
When both the momentum conservation and pseudospin are physically legitimate,
$b \simeq 0$.
When we can only rely on the pseudospin, 
we expect $b/a$ to be 1/3 by using the ratio of 
$\oint \cos^2\Theta \sin^2\Theta d\Theta$ ($=\pi/4$) to 
$\oint \cos^4\Theta d\Theta$ ($=3\pi/4$).
Figure~\ref{fig:polarp} shows the polar plots for these two cases.
If the residual $D$ band intensity decreases by putting a polarizer
for the scattered light, such a signal would evidence the effect of the pseudospin.
The $\Theta_{\rm in}$ independent $c$ term may originate from a point
defect or on-site deformation potential (see Appendix~\ref{app:kekule}) and
obscures the $D$ band polarization dependence. 

\subsection{Kohn anomaly and transport}\label{ssec:armka}

In ANRs, because of the conical shape of the energy spectrum, 
momentum conservation tells us that a vertical (direct) electron-hole
pair can couple to the $\Gamma$ point phonon. 
Furthermore, due to the BZ holding, the direct electron-hole pair can
also couple to the $D$ band phonons.
As a result, the G and $D$ bands in Raman spectra 
can exhibit a similar Kohn anomaly effect.
The probability amplitude for the process whereby
the $D$ and $G_y$ bands change into 
a vertical electron-hole pair is given by 
using $| \phi^{v}_n \rangle = \sigma_z | \phi^{c}_n \rangle$,
$\sigma_x\sigma_z=-i\sigma_y$, Eq.~(\ref{eq:epD}), and Eq.~(\ref{eq:uyg}), as
\begin{align}
\begin{split}
 & \langle \phi^{c}_n | H_{\rm off}({\bu}_{q \simeq 2\pi/3})  | \phi^{v}_n \rangle
 = -3i \langle \sigma_y \rangle^{cc}_{nn}, \\
 & \langle \phi^{c}_n | H_{\rm off}(\bu_{0})  | \phi^{v}_n \rangle
 = 3i \langle \sigma_y \rangle^{cc}_{nn}.
\end{split} 
 \label{eq:ka}
\end{align}
Since $\langle \sigma_y \rangle^{cc}_{nn}=\sin\Theta_n$,
electrons at $\Theta_n=\pi/2$ or $-\pi/2$ contribute to the 
Kohn anomaly.
Equation~(\ref{eq:ka}) clearly shows that 
the behavior of the Kohn anomaly for the $D$ band is correlated to 
that for the $G_y$ band:
if the $D$ band exhibits hardening/softening
due to the Kohn anomaly effect, the $G_y$ band also exhibits
hardening/softening.
Recently, Zhang and Li~\cite{zhang11}
have observed a Kohn anomaly effect for the G band at an armchair graphene edge. 
Interestingly they also found a correlation between the 2$D$ band and the $G$ band.

It is interesting to note that
the pseudospin for electron-hole pair creation 
is identical to the matrix elements 
of the velocity operator along the armchair edge,
which is written as
$\langle \phi^{c}_m |v_y | \phi^{c}_n \rangle=\delta_{mn}v_{\rm
F}\langle \sigma_y \rangle_{mn}^{cc}$.~\cite{sasaki11-dc}
Thus, the Kohn anomaly effect is fundamentally 
related to the velocity or current behavior.
If the velocity is suppressed, 
then the Kohn anomaly effect is suppressed as well.
The velocity perpendicular to the armchair edge $v_x$ 
is suppressed by electron reflection, while 
the velocity along the edge $v_y$ is not.
As a result, only the $D$ and $G_y$ bands 
can experience a strong Kohn anomaly effect.~\cite{sasaki10_pola}
In the presence of impurities,
the electronic velocity (current) along the edge (ribbon)
might also be suppressed due to the scattering.
However, in metallic ANR, 
because Berry's phase can protect 
the electronic current along the armchair edge 
from decaying,~\cite{sasaki10-forward}
the Kohn anomaly effect for the $D$ and $G_y$ bands
should be robust against impurity potentials.
We consider that the $D$ and $G_y$ bands
undergo a strong Kohn anomaly effect 
even when we take account of the effect of impurity potentials.

\section{Discussion}\label{sec:dis}

Here we discuss the possible factors that 
obscure the double-component feature of the $D$ band
proposed in Sec.~\ref{ssec:tp}.
First, the resonance effect suppresses the intensity of one of the two
components.
Suppose that the electron at $k_0$ is resonant with 
the incident laser energy.
Then the Raman intensity for the process in which 
the photo-excited electron at $k_0$ emits a
phonon with momentum $q_0$ is enhanced as compared with 
the Raman intensity involving a phonon with momentum $q_\pi$.
Such a resonance effect is apparent in thin ANRs, 
whereas it is negligible for thick ANRs 
if the ANR width (W) exceeds a critical value.
The critical $W$ is estimated from the condition whereby
the difference between $k_0-2\pi/3$ and
$2\pi/3-k_\pi$ caused by the trigonal warping effect
[Eq.~(\ref{eq:thetae})]
is larger than the spacing of each sub-band, $\pi/(N+1)$.
This condition leads to $W > 8(\gamma/E_L)^2$ nm:
$W > 13$ nm when $E_L=2.3$ eV.
Second, 
for thick ANRs satisfying $W > 8(\gamma/E_L)^2$ nm or 
for armchair edges of graphene, 
the energy gap is smaller than the energy of the $D$ band.
As a result, the width of each peak can be wide enough to obscure the
splitting structure since the $D$ band undergoes a strong Kohn anomaly
effect (Sec.~\ref{ssec:armka}).
To suppress the broadening of each peak caused by 
the Kohn anomaly effect, 
the Fermi energy needs to be placed away from the Dirac point ($|E_{\rm
F}|> 0.1$ eV).
Third, 
the parallel configuration is the most suitable polarization setting
with which to observe the splitting of the $D$ band 
because the probability $\cos^6\Theta$ exhibits prominent peaks at
$\Theta=0$ or $\pi$. 
To observe the splitting of the $D$ band, 
it is important to recognize that 
the results obtained in Secs.~\ref{ssec:tp}, \ref{ssec:lpd}, and
\ref{ssec:armka} are closely correlated.
Among the conclusions obtained in Secs.~\ref{ssec:tp}, \ref{ssec:lpd}, and
\ref{ssec:armka}, 
we think that the light polarization dependence of the $D$ band Raman intensity
is the most direct conclusion derived from the pseudospins of 
the electron-phonon and optical matrix elements. 
Indeed, many studies have reported the $D$ band polarization
behavior.~\cite{canifmmode04sec,you08,gupta09,casiraghi09,cong10,barros11}
However, there is a small possibility that only the pseudospin of the
optical matrix element is the result of the observed polarization dependence
for the $D$ band if there is a strong resonance effect.~\footnote{
At the graphene edge, it is difficult to expect a resonance effect, 
such as the Van Hove singularity for electrons and phonons.
The polarization dependence of the $D$ band at a graphene edge supports
the pseudospin dependence of the electron-phonon matrix element.}
A strong piece of evidence for the pseudospin of the electron-phonon
matrix element is provided by the splitting of the $D$ band.

A multicomponent $D$ band has been seen in several sets of Raman data
obtained at the graphene edge.
Ferrari {\it et al.}~\cite{ferrari06}
and Gupta {\it et al.}~\cite{gupta09} 
reported that for graphite,
the $D$ band can be well fitted by a doublet with two broad components, 
although a single-layer graphene edge produces a narrow single-component
$D$ band.
Our theory relates to the armchair edge of a single-layer graphene.
However, the results that we have obtained for ANRs
may be applicable to single-wall carbon nanotubes because
defects in carbon nanotubes result in the formation of a standing wave.
Suzuki and Hibino~\cite{suzuki11} 
reported that there is a correlation between
the behaviors of the $D$ and $G$ bands under doping.
Moreover, they observed $D$ band splitting
when the excitation laser energy was at its largest (532 nm), while 
such a splitting was not seen when they used lower excitation energies,
633 nm and 785 nm. 
These observations are consistent with our results, that is,
Kohn anomaly effects for the $D$ and $G$ bands are correlated and 
the $D$ band splitting increases with increasing $E_L$. 
We note that work has already been published on the splitting
of the $D$ band in carbon nanotubes,~\cite{maultzsch01,zolyomi03} 
in which the authors propose a completely different mechanism from ours
(trigonal warping effect for phonon dispersion).

\section{Conclusion}\label{sec:con}

By constructing the matrix element of the electron-phonon interaction
for the $D$ band analytically, we have shown that 
the $D$ band in an ANR is represented as a first order, single resonance
process in a reduced BZ, as well as the $G$ band.
This explains clearly and naturally the fact that the $D$ and $G$ bands
show a similar intensity in many experiments for armchair edge.
The matrix element is proportional to the pseudospin 
$\langle \sigma_x \rangle_{nn}^{cc}$, and the $D$ band couples
selectively to the two pseudospin states $\langle \sigma_x
\rangle_{nn}^{cc}=\pm1$ corresponding to the bonding and antibonding orbitals.
The pseudospin is the origin of the branches of phenomena associated with
the $D$ band in ANRs, as shown in Secs.~\ref{ssec:tp}, \ref{ssec:lpd}, and
\ref{ssec:armka}.
The polarization dependence of the $D$ band 
is consistent with the previous experimental reports.
Direct evidence for the pseudospin of the $D$ band is provided by 
the $D$ band splitting at the armchair edge, which has not been previously
reported.

\section*{acknowledgments}

K.S. is supported by 
a Grant-in-Aid for Specially Promoted Research
(Grant No.~23310083) from the Ministry of Education, Culture, Sports, Science
and Technology.

\appendix

\section{Generalized Kekul\'e distortion}\label{app:kekule}

When the amplitudes of the acoustic and optical modes
are not exactly the same, we may consider the extension
of Eq.~(\ref{eq:mode}) to the general case as 
\begin{align}
 (1-\epsilon) \sin(q x) 
 \begin{pmatrix}
  {\bf e}_x \cr {\bf e}_x
 \end{pmatrix}
 + (1+\epsilon) \cos(q x) 
 \begin{pmatrix}
  {\bf e}_y \cr -{\bf e}_y
 \end{pmatrix}.
\end{align}
In this Appendix, we show how 
the analysis is affected by the deviation $\epsilon$.

The generalized displacement vector is written as 
${\bu}_{q}(x)+\epsilon {\bu}_{-q}(x)$, and 
the corresponding electron-phonon matrix elements 
are 
\begin{align}
 \langle \phi^{s'}_m| H_{\rm on/off}({\bu}_q+\epsilon {\bu}_{-q})) |
 \phi^{s}_n \rangle 
 =\langle \phi^{s'}_m| H_{\rm on/off}({\bu}_q) | \phi^{s}_n \rangle 
 + \epsilon \langle \phi^{s'}_m| H_{\rm on/off}({\bu}_{-q}) | \phi^{s}_n \rangle.
\end{align}
Because the last term is given by 
replacing $q$ with $-q$ in 
Eqs.~(\ref{eq:Mu}) and (\ref{eq:Mu_on}),
we see that 
the off-site component is suppressed for $q \simeq 2\pi/3$, 
$\epsilon \langle \phi^{s'}_m| H_{\rm off}({\bu}_{-q\simeq -\frac{2\pi}{3}}) |
 \phi^{s}_n \rangle \simeq 0$,
and that the correction to the matrix element of 
Eq.~(\ref{eq:epD}) 
arises only from the
on-site component,
$\epsilon \langle \phi^{s'}_m| H_{\rm on}({\bu}_{-q\simeq -\frac{2\pi}{3}}) |
 \phi^{s}_n \rangle
 \simeq -3\epsilon \langle \sigma_0 \rangle_{mn}^{s's}$.
Thus, the total electron-phonon matrix element for the deformed Kekul\'e
distortion is given by
\begin{align}
 3 g_{\rm off}\langle \sigma_x \rangle_{mn}^{s's} 
 -3\epsilon g_{\rm on}\langle \sigma_0 \rangle_{mn}^{s's},
\label{app:Melph}
\end{align}
where $g_{\rm on}$ ($g_{\rm off}$) denotes the coupling constant
of the on-site (off-site) deformation potential.
Since the intensity is relevant to the square of the matrix element,
the correction can include the crossing term 
$-18\epsilon g_{\rm on}g_{\rm off}\langle \sigma_x \rangle_{nn}^{ss} \langle \sigma_0 \rangle_{nn}^{ss}$
and the order $\epsilon^2$ term $9\epsilon^2 g^2_{\rm on}(\langle \sigma_0 \rangle_{nn}^{ss})^2$.
The crossing term does not contribute to the intensity because it
vanishes after the integral over $\Theta_n$.
Thus, the leading contribution is given by the order $\epsilon^2$.
The effect of the deviation is negligible when 
$|\epsilon|  \ll |g_{\rm off}/g_{\rm on}|$, where $|g_{\rm off}/g_{\rm
on}|\sim 1/3$ was suggested by the result of density functional theory.~\cite{porezag95}
Note also that the pseudospin is proportional to $\sigma_0$, which shows that
the deviation is not relevant to the polarization dependence of the $D$
band intensity but relevant to the $c$ parameter in Eq.~(\ref{eq:ID}) 
or residual $D$ band intensity.

Using Eq.~(\ref{app:Melph}),
a quick comparison can be made 
between the predictions of our
model on which the $D$ band is described as coming from a Kekul\'e
distortion and other models where the $D$ band is described solely as
a combination of intervalley optical phonons. 
The consequence of other models may be obtained by 
setting $\epsilon=1$ in Eq.~(\ref{app:Melph}).
Then, the matrix element is dominated by the on-site component and 
the pseudospin $\sigma_x$ is suppressed. 
As a result, the $D$ band does not exhibit the polarization dependence.

\section{Derivations of Eqs.~(\ref{eq:Mu}) and (\ref{eq:Mu_on}), and
 boundary condition for phonon}\label{app:ep}

Below we give the derivation of the electron-phonon matrix elements.
Note that 
$\bu_q(J)$ in Eq.~(\ref{eq:umodes})
consists only of the optical mode at the fictitious edge site of $J=0$.
This means that Eq.~(\ref{eq:umodes}) results from 
a specific choice of phonon boundary condition.
To make the derivation general enough to cover
the possible boundary conditions,
we first generalize $\bu_q(J)$ of Eq.~(\ref{eq:umodes})
by inserting a phase shift of $\varphi$ as
\begin{align}
 \sin(qJ+\varphi) 
 \begin{pmatrix}
  {\bf e}_x \cr {\bf e}_x
 \end{pmatrix}
 + \cos(qJ+\varphi) 
 \begin{pmatrix}
  {\bf e}_y \cr -{\bf e}_y
 \end{pmatrix}.
 \label{app:u}
\end{align}
For $\varphi=0$, Eq.~(\ref{app:u}) reproduces Eq.~(\ref{eq:umodes}).
For $\varphi=\pi/2$, Eq.~(\ref{app:u}) leads to
\begin{align}
 \bar{\bu}_{q}(J) = 
 \cos(qJ)
  \begin{pmatrix}
  {\bf e}_x \cr {\bf e}_x
 \end{pmatrix}
 - \sin(qJ)
 \begin{pmatrix}
  {\bf e}_y \cr -{\bf e}_y
 \end{pmatrix}.
\end{align}
This $\bar{\bu}_q(J)$
consists only of the acoustic mode at $J=0$, which is contrasted with
that $\bu_q(J)$ of Eq.~(\ref{eq:umodes}) that 
consists only of the optical mode at $J=0$.~\footnote{
The radial-breathing like mode described by Zhou and Dong~\cite{zhou07apl}
corresponds to the acoustic mode with the form $\cos(qJ) ({\bf 
e}_x,{\bf e}_x)$ with $q=\pi/(N+1)$.}
The phase $\varphi$ is not very meaningful for a periodic system
without an edge, such as a nanotube,
however, for a nanoribbon with an edge, phase $\varphi$
should be taken into account
because it is possible that the phonon boundary condition
could be sensitive to the situation of the armchair edge.
It turns out that $\varphi$ changes the momentum conservation slightly,
while it does not modify the pseudospin structure of the matrix element.
Thus, an analysis based on the pseudospin seems plausible, while 
the results based only on the momentum conservation might be physically
fragile.  
Similarly, the $\bv_q(J)$ mode can be generalized as
\begin{align}
 \cos(qJ+\varphi) 
 \begin{pmatrix}
  {\bf e}_x \cr -{\bf e}_x
 \end{pmatrix}
 - \sin(qJ+\varphi) 
 \begin{pmatrix}
  {\bf e}_y \cr {\bf e}_y
 \end{pmatrix}.
\end{align}
In addition, the physics of graphene edge is, in some sense, an
understanding of the boundary conditions at edges.
The importance of the boundary conditions for electrons
has been widely recognized, while those for the phonon mode and for
electron-phonon interactions are yet to be understood.

The off-site electron-phonon matrix elements 
for the displacement vector
${\bu}_q(J)$ of Eq.~(\ref{app:u})
are written as
\begin{align}
 \langle \phi^{s'}_m| H_{\rm off}({\bu}_q) | \phi^s_n \rangle
 = \sum_{J=1}^N \left[ \phi^{s'}_{m,J} \right]^\dagger
 \left(
 \delta h^{+}_J G^+ \phi^s_{n,J+1} + 
 \delta h_{J} \phi^s_{n,J} +
 \delta h_{J}^- G^- \phi^s_{n,J-1} \right).
 \label{app:Hep}
\end{align}
On the right-hand side,
we have defined the translational operators,
\begin{align}
 G^+ \equiv 
 \begin{pmatrix}
  e^{ik_yb} & 0 \cr 0 & 1
 \end{pmatrix},
 \ \
 G^- \equiv
 \begin{pmatrix}
  1 & 0 \cr 0 & e^{-ik_yb}
 \end{pmatrix},
\label{app:defG}
\end{align} 
and the 2$\times$2 matrices of the deformation potential as
\begin{align}
\begin{split}
 & \delta h_{J} = 
 \begin{pmatrix}
  0 & \delta \gamma_{1}(J) \cr
  \delta \gamma_1(J) & 0
 \end{pmatrix}, 
 \\  
 & \delta h_{J}^+ = 
 \begin{pmatrix}
  0 & \delta \gamma_3(J) \cr
  \delta \gamma_2(J+1) & 0 
 \end{pmatrix},
 \\
 & \delta h_{J}^- =
 \begin{pmatrix}
  0 & \delta \gamma_2(J) \cr
  \gamma_3(J-1) & 0
 \end{pmatrix},
\end{split} 
\label{app:hjuapp}
\end{align}
where 
\begin{align}
\begin{split}
 & \delta \gamma_1(J) = 
 \left[ {\bu}_q^{\rm B}(J) - {\bu}_q^{\rm A}(J) \right]\cdot {\bf e}_1, \\
 & \delta \gamma_2(J) =
 \left[ {\bu}_q^{\rm B}(J-1) - {\bu}_q^{\rm A}(J) \right]\cdot {\bf e}_2, \\
 & \delta \gamma_3(J) =
 \left[ {\bu}_q^{\rm B}(J+1) - {\bu}_q^{\rm A}(J) \right]\cdot {\bf e}_3,
\end{split}
 \label{app:deltag}
\end{align}
with ${\bu}_q^{\rm A}(J) = \sin(qJ+\varphi){\bf
e}_x+\cos(qJ+\varphi){\bf e}_y$
and 
${\bu}_q^{\rm B}(J)=\sin(qJ+\varphi){\bf
e}_x-\cos(qJ+\varphi){\bf e}_y$ from \ref{app:u}.
In Eq.~(\ref{app:deltag}) ${\bf e}_a$ ($a=1,2,3$) are
the dimensionless unit vectors pointing from an A-atom to the
nearest-neighbor B-atoms:
\begin{align}
 {\bf e}_1={\bf e}_y,
 {\bf e}_2=- \frac{\sqrt{3}}{2} {\bf e}_x - \frac{1}{2} {\bf e}_y,
 {\bf e}_3=\frac{\sqrt{3}}{2} {\bf e}_x - \frac{1}{2} {\bf e}_y. 
\label{app:ea}
\end{align}
By inserting Eq.~(\ref{app:u}) into Eq.~(\ref{app:deltag}),
we rewrite Eq.~(\ref{app:hjuapp}) as
\begin{align}
\begin{split}
 & \delta h_{J} = -2 \sigma_x \cos(qJ+\varphi),\\
 & \delta h_{J}^+ =  2 \sigma_x
 \cos\left(qJ+\varphi+\frac{q}{2} \right)
 {\bf e}_q \cdot {\bf e}_3, \\
 & \delta h_{J}^- =  2 \sigma_x
 \cos\left(qJ+\varphi-\frac{q}{2} \right) 
 {\bf e}_q \cdot {\bf e}_3,
\end{split}
\label{app:hj}
\end{align}
where we have defined
\begin{align}
 {\bf e}_q = \sin \left(\frac{q}{2}\right) {\bf e}_x - 
 \cos \left( \frac{q}{2} \right) {\bf e}_y.
\end{align}
It is easy to show that ${\bf e}_q$ satisfies
\begin{align}
 {\bf e}_q \cdot {\bf e}_3 = 
 \sin\left(\frac{q}{2}+\frac{\pi}{6} \right) = 
 {\bf e}_{-q} \cdot {\bf e}_2.
\end{align}
Finally, by putting Eq.~(\ref{app:hj}) into Eq.~(\ref{app:Hep}),
we obtain
\begin{footnotesize}
\begin{align}
 & \langle \phi^{s'}_m| H_{\rm off}({\bu}_q) | \phi^s_n \rangle \nn \\
 =& \sum_{J=1}^N \left[ \phi^{s'}_{m,J} \right]^\dagger
 \left\{ 2T{\bf e}_q \cdot {\bf e}_3  \cos\left(qJ+\varphi +\frac{q}{2} \right)
 \phi^s_{n,J+1} - 2\sigma_x \cos\left(qJ+\varphi \right) \phi^s_{n,J} +
 2T^{-1}{\bf e}_q \cdot {\bf e}_3  \cos\left(qJ+\varphi-\frac{q}{2} \right)
 \phi^s_{n,J-1} \right\} \nn \\
 =& \sum_{J=1}^N \left[ \phi^{s'}_{m,J} \right]^\dagger
 \left\{ -2\cos(qJ+\varphi) \left[ \sigma_x+
 K_n^s{\bf e}_q \cdot {\bf e}_3  \cos\left(\frac{q}{2}
 \right) 
 \right]  \phi^s_{n,J}
 -2 \sin(qJ+\varphi) {\bf e}_q \cdot {\bf e}_3 
 \sin\left(\frac{q}{2} \right) 
 \left[ 
 T \phi^s_{n,J+1} -T^{-1} \phi^s_{n,J-1}
 \right] 
 \right\} \nn \\
 =& \Bigg\{ -
 \left( \frac{2}{N} \sum_{J=1}^N \sin(k_m J)\cos(qJ+\varphi)
 \sin(k_n J)\right)
 \left[ 2 + 2{\bf e}_q \cdot {\bf e}_3  \cos\left(\frac{q}{2}\right)
 \right] \nn \\
 & \ \ \ - \left( \frac{2}{N} \sum_{J=1}^N 
 \sin(k_m J) \sin(qJ+\varphi) \cos(k_n J) \right)
 \left[ 4 {\bf e}_q \cdot {\bf e}_3 \sin\left(\frac{q}{2}
 \right) \sin (k_n) \right] \Bigg\}
 \langle \sigma_x  \rangle^{s's}_{mn},
 \label{app:matu}
\end{align}
\end{footnotesize}
where
the matrix $T$ ($T^{-1}$) is defined by 
$T \equiv \sigma_x G^+$ ($T^{-1} \equiv \sigma_x G^-$).
Note that 
the energy eigen equation 
for the electron in an ANR
is written in terms of these $T$, $T^{-1}$, 
and $K^s_n \equiv \sigma_x + (\varepsilon^s_{k_n,k_y}/\gamma) \sigma_0$ as~\cite{sasaki11-armwf}
\begin{align}
 T \phi^s_{n,J+1} + K^s_n \phi^s_{n,J} + T^{-1} \phi^s_{n,J-1} = 0.
\end{align}
The energy eigen equation has been used 
to obtain the third line in Eq.~(\ref{app:matu}).
Note also that 
the small term, $\varepsilon^s_{k_n,k_y}/\gamma$, has been omitted
to obtain the last line.
By using Eq.~(\ref{eq:bw}), we obtain the following equation,
\begin{align}
 T \phi^s_{n,J+1} -T^{-1} \phi^s_{n,J-1} = 
  2 \sin(q) Te^{-ik_y} \frac{1}{\sqrt{N}}e^{-ik_y(J-1)} \cos(k_n J) 
 \begin{pmatrix}
  e^{-i\Theta_n} \cr s 
 \end{pmatrix},
\end{align}
which has been used to get the last line.
We reproduce Eq.~(\ref{eq:Mu}) by putting $\varphi=0$ in Eq.~(\ref{app:matu}).
When $\varphi=\pi/2$, the momentum conservation in Eq.~(\ref{app:matu})
changes slightly.
It is important to mention that the pseudospin is independent of the
choice of the value of $\varphi$.
Because the matrix element for the generalized $\bv_q(J)$
can be obtained by replacing $\bu_q(J)$ with $\bv_q(J)$
in the above calculation, we omit the derivation 
of the matrix elements for $\bv_q(J)$.

The on-site deformation potential is written as
\begin{align}
 \delta h_J = 
 \begin{pmatrix}
  \delta h^{\rm A}_J & 0 \cr
  0 & \delta h^{\rm B}_J
 \end{pmatrix},
\end{align}
where
\begin{align}
\begin{split}
 \delta h^{\rm A}_J 
 &= {\bu}^{\rm B}_q(J) \cdot {\bf e}_1
 + {\bu}^{\rm B}_q(J-1) \cdot {\bf e}_2
 + {\bu}^{\rm B}_q(J+1) \cdot {\bf e}_3,
 \\
 \delta h^{\rm B}_J 
 &= - \left\{
 {\bu}^{\rm A}_q(J) \cdot {\bf e}_1
 + {\bf u}^{\rm A}_q(J-1) \cdot {\bf e}_3
 + {\bf u}^{\rm A}_q(J+1) \cdot {\bf e}_2 \right\}.
\end{split}
 \label{app:hjon}
\end{align}
The matrix element is obtained by inserting the above $\delta h_J$
into Eq.~(\ref{app:Hep}) and setting $\delta h_J^\pm=0$.
By putting Eq.~(\ref{app:u}) into Eq.~(\ref{app:hjon}),
we obtain for $J=2,\ldots,N-1$ that
\begin{align}
 \delta h_J
 = 4\sigma_0 \cos(qJ+\varphi)
 \sin\frac{q}{2} \cos\left(\frac{\pi}{6}+\frac{q}{2} \right),
 \label{eq:honj}
\end{align}
which reproduces Eq.~(\ref{eq:Mu_on}) when $\varphi=0$.
The calculation of 
the on-site deformation potential at the boundary,
$\delta h_1$ and $\delta h_{N}$,
requires careful treatment
because there is no site at $J=0$ and $J=N+1$.
In general, we have a boundary deformation potential,
and we cannot use Eq.~(\ref{eq:honj}) to estimate
the boundary deformation potential.
This fact can be understood in Fig.~\ref{fig:dis}(b)
by looking at the relative displacement 
of the A (B) atom at the boundary site with $J=1$
in relation to that of the nearest-neighbor B (A) atom at $J=2$.
For example, we have $\delta h_1({\bu}_{q})= {\rm diag}(1/2,1/2)$
even if $q=0$.
This is not expected from a simple application of Eq.~(\ref{eq:honj})
to the boundary site at $J=1$.
The effect of the boundary deformation potential on the electronic state
may prove important when we consider localized states.


\end{document}